\documentclass[twocolumn]{aastex631}

\usepackage{graphicx}	
\usepackage{amsmath}	
\usepackage{amssymb}	
\usepackage{makecell}
\usepackage{comment}
\usepackage{multirow}

\begin{document}

\title{Radio Emission from Broad Absorption Line Quasars}


\correspondingauthor{Sina Chen}
\email{sina.chen@campus.technion.ac.il}

\author{Sina Chen}
\affiliation{Physics Department, Technion, Haifa 32000, Israel}

\author{Ehud Behar}
\affiliation{Physics Department, Technion, Haifa 32000, Israel}

\author{Ari Laor}
\affiliation{Physics Department, Technion, Haifa 32000, Israel}

\author{Nahum Arav}
\affiliation{Department of Physics, Virginia Tech, Blacksburg, VA 24061, USA}

\begin{abstract}

Broad Absorption Line Quasars (BALQs) generally exhibit significant outflows that may interact with the surrounding medium, resulting in radio emission.
We selected a sample of 13 powerful radio-quiet (RQ) BALQs, where the UV outflow kinetic power is measurable, and detected nine of them with the Very Large Array A configuration at 5.5\,GHz and 9.0\,GHz.
The radio emission is mostly unresolved and is generally constrained within a scale of $<$ 1--4\,kpc.
In the nine detected objects, the radio spectral slope $\alpha_{5.5-9.0}$ is steep ($< -0.5$) in five objects and is flat or inverted ($> -0.5$) in four objects.
We discuss how the steep-slope emission can be associated with the UV outflows, and how the flat-slope emission can be intrinsically steep but flattened by free-free absorption from the UV outflowing gas.
However, we find no correlation between the radio luminosity and the estimated outflow kinetic power, which suggests that the outflows are likely not a major source of the observed radio emission.
In addition, the radio loudness of these RQ BALQs is comparable to that of typical RQ quasars, implying that the UV outflows likely do not produce stronger radio emission compared to non-BALQs.
Follow-up radio observations can test the free-free absorption interpretation and can be used as a new probe for outflows in AGN.

\end{abstract}


\section{Introduction}

Active Galactic Nuclei (AGN) outflows/winds are often invoked to affect the formation and evolution of galaxies over cosmological times \citep{Silk1998, DiMatteo2005}, and as we observe them today.
The effect of outflows on galaxies and their star formation is hard to measure directly, and is thus poorly understood.
They are often quantified for simplicity in terms of relative energy feedback, which is the ratio of the kinetic power $\dot{E}_{\rm k} = \frac{1}{2} \dot{M}_{\rm out} v^2$ to the AGN bolometric luminosity $L_{\rm bol}$, where $\dot{M}_{\rm out}$ is the mass outflow rate and $v$ is the outflow velocity.
Cosmological simulations suggest that the relative energy feedback of a few percent is sufficient to affect the galaxy \citep{DiMatteo2005, Hopkins2010}. However, direct measurements of $\dot{M}_{\rm out}$ have proven to be elusive.

Broad Absorption Line Quasars (BALQs) are one manifestation of such outflows.
The absorption lines tend to be blueshifted relative to the emission lines and their rest-frame wavelengths, indicating the presence of winds emanating from the AGN center with velocities as large as $v \approx 0.2\,c$ \citep{Ganguly2008}.
One promising and proven method to measure the location and thus to estimate the mechanical energy of these winds is by spectroscopic density diagnostics, which measure the wind distance from the AGN center through the ionization parameter and the gas density $U \propto n^{-1}r^{-2}$, and thus the $\dot{M}_{\rm out}$ with an assumed wind geometry \citep{Arav2018}.
Such methods have been applied to a sample of BALQs and mini-BALQs, which show absorption troughs with the widths of $\Delta v >$ 2000~km\,s$^{-1}$ and 500~km\,s$^{-1}$ $< \Delta v <$ 2000~km\,s$^{-1}$, respectively, in their spectra \citep{Miller2020,Byun2022}.
These measurements revealed highly extended non-relativistic winds, from $\sim$ 0.1\,pc up to $\sim$ 70\,kpc away from the nucleus, indeed reaching the required relative energy feedback of a few percent.

Measuring the mechanical energy of an outflow alone is not yet a proof of AGN feedback.
In order for the outflow to impact the galaxy, it needs to interact with the ambient medium.
One form of impact is by generating shocks that accelerate particles to relativistic energies \citep{Jiang2010}.
The interaction of these particles with the magnetic field produces optically thin synchrotron emission that can be observed in radio.
In fact, the correlation of the 5\,GHz flux with the [O\,III] line width in a sample of AGN has been suggested to be the evidence for such shocked outflows.
This could be the sought after feedback process, observed in both local \citep{Zakamska2014} and high-redshift \citep[$z \sim 3$;][]{Hwang2018} AGN.

AGN are typically classified as either radio-loud (RL) or radio-quiet (RQ) based on the radio loudness $R$, which is the 5\,GHz to 4400\,{\AA} flux ratio.
RL sources have a ratio larger than 10 and RQ ones smaller than 10 \citep{Kellermann1989}.
In RL AGN, the radio emission is primarily produced by a powerful relativistic jet, while the physical origin of the radio emission in RQ AGN is still under debate.
Various mechanisms have been proposed, including a low-power jet, an AGN-driven wind, the accretion disk corona, star formation (SF), and free-free emission \citep[see][for a review]{Panessa2019}.

Like the AGN population at large, the BALQs are mostly RQ \citep{Stocke1992, Shankar2008, Morabito2019}.
A study of the absorption line profiles of the radio-detected BALQs suggests an intrinsic connection between the wind and the radio emission, with the radio emission coming from the wind itself \citep{Petley2022}.
If this is the case, it is expected that the radio emission is optically thin with a steep spectral slope $\alpha < -0.5$.
In a sample of RQ Palomar–Green (PG) quasars, optically thin synchrotron sources are shown to have high Eddington ratios $L/L_{\rm Edd} > 0.3$ \citep{Laor2019}.
This suggests that the radio emission can be associated with a radiatively driven wind.

The present work aims to look for the wind associated radio emission, which is direct evidence for the AGN wind interacting with the ambient medium.
For this purpose, we employed a sample of BALQs and mini-BALQs, whose $\dot{E}_{\rm k}$ has been measured, and observed them at 5.5 and 9.0\,GHz with the Very Large Array (VLA) A configuration.
The paper is organized as follows. The sample selection, data reduction, and data analysis are described in Sections 2, 3, and 4, respectively. The results can be found in Section 5, and their discussion in Section 6. The summary is given in Section 7.
Throughout this work, we use the flux density and spectral index convention of $F_\nu \propto \nu^{\alpha}$.
We adopt a standard $\Lambda$CDM cosmology with a Hubble constant $H_0$ = 70\,km\,s$^{-1}$\,Mpc$^{-1}$, $\Omega_{\Lambda}$ = 0.73 and $\Omega_{\rm M}$ = 0.27 \citep{Komatsu2011}.

\section{Sample Selection}

We focus on BALQs ($\Delta v >$ 2000~km\,s$^{-1}$) and mini-BALQs (500~km\,s$^{-1}$ $< \Delta v <$ 2000~km\,s$^{-1}$), where the width of the absorption troughs ($\Delta v >$ 500~km\,s$^{-1}$) assures that they arise from outflows connected with the quasars.
About $\sim$ 90\% of all BALQs and mini-BALQs show absorption troughs only from high ionization species, such as \ion{C}{4} and \ion{Si}{4} \citep{Trump2006}.
We therefore concentrate on the high-ionization BALQs in order for our sample to be representative of the majority of BALQs and mini-BALQs.

We select all such objects published in the literature where the outflow distances from the central source $r_{\rm out}$ and their $\dot{E}_{\rm k}$ are reported.
There are two such groups including 16 objects.
The first one is from the ground-based observations targeting the \ion{S}{4}\,($\lambda$1062.66~{\AA}) and \ion{S}{4}*\,($\lambda$1072.96~{\AA}) troughs \citep{Xu2019}.
The second one is from the Hubble Space Telescope (HST) observations targeting the rest-frame extreme ultraviolet (EUV) at 500--1000~{\AA} \citep{Arav2020}.
The assumptions and methodology used to derive $r_{\rm out}$, $\dot{E}_{\rm k}$, and their assigned uncertainties are described in the two papers mentioned above, and all the parameters are summarized in \citet[][Table A2]{Miller2020}.
We make one exception to include an object (J0242+0049) whose outflows were measured using troughs from low ionization species, specifically \ion{Fe}{2} \citep{Byun2022}.
We do so since one of its outflows has the largest $r_{\rm out}$ ($\approx$ 67~kpc) reported to date.

The sample spans a redshift range of $z \sim 0.4-2.8$, overlapping with the era of peak star formation and AGN activity.
All the quasars were selected to be bright with a minimum continuum flux of $2 \times 10^{-15}$~erg\,s$^{-1}$\,cm$^{-2}$\,{\AA}$^{-1}$ at the rest-frame 500--1000~{\AA}.
This selection of the flux and the redshift naturally results in objects with high luminosities and correspondingly large black hole (BH) masses $M_{\rm BH}$.
The sample is representative of high-ionization outflows, which are the majority of the observed quasar outflows.

The objects have been reported to have outflows extending from $\sim$ 0.1\,pc up to $\sim$ 70\,kpc with large uncertainties, of which seven show multi-velocity components in the absorption lines.
The total $\dot{E}_{\rm k}$ of all the velocity components are in the ranges of $10^{42.4}-10^{46.9}$~erg\,s$^{-1}$, which are about $10^{-5}-1~L_{\rm bol}$, with uncertainties of about $\pm$0.1--0.9\,dex, which are 1$\sigma$ errors (68\% confidence values) in statistics.
The procedure of measuring the outflow kinetic powers from the quasar UV spectra is detailed in \cite{Borguet2012}.
There are additional systematic errors not included due to the assumptions used:
(a) the geometry of the outflowing material is assumed to be in the form of a thin, that is the thickness less than half of the radius, and partially filled shell \citep{Xu2018};
(b) the kinetic power is proportional to the solid angle of the outflow around the quasar, which is canonically assumed to be the fraction of all quasars showing these types of absorption outflows \citep{Arav2013};
(c) the calculation assumes solar abundances, as well as a specific spectral energy distribution for the quasar emission.
Plausible deviation from these assumptions can introduce a $\pm$0.7\,dex scatter in the reported values \citep{Xu2018}.

The deduced outflow $\dot{E}_{\rm k}$ are not subjected to the time-dependent ionizing luminosity \citep[e.g.][]{Arav2015}, which is the reason of the changes in the photoionization equilibrium of the absorbing gas in most of the absorption line variability.
A comparison of $\dot{E}_{\rm k}$ and $L_{\rm bol}$ between our sample and a sample from \cite{Fiore2017}, where the entire collection consists of over 80 AGN and includes ionized and molecular emission outflows in addition to UV and X-ray absorption outflows, is present in \citet[][Figure 1]{Miller2020}.
The $\dot{E}_{\rm k}$ range of our sample overlaps that of \cite{Fiore2017} for the above $L_{\rm bol}$ range.

Among these 17 objects, five were detected at 1.4~GHz in the Faint Images of the Radio Sky at Twenty-Centimeters \citep[FIRST;][]{Helfand2015} with a resolution of 5\arcsec, of which four with $R \gtrsim 10$ (7--51) were excluded from the observations in order to avoid confusion with the strong jet emission.
(An estimate of the $R$ value is described in Section 4.)
We are left with 13 RQ objects in the sample, including seven BALQs and six mini-BALQs.
The values of $L_{\rm bol}$, $M_{\rm BH}$, $L/L_{\rm Edd}$, $\dot{E}_{\rm k}$, and $r_{\rm out}$ of the sample from literature are summarized in Table~\ref{sample}.

The $L_{\rm bol}$ and $M_{\rm BH}$ of 12 of the 13 objects are obtained from a quasar catalog \citep{Vestergaard2006, Vestergaard2009, Shen2011} based on the Sloan Digital Sky Survey (SDSS) Data Release 7.
We use the $M_{\rm BH}$ based on the H$\beta$ emission line when it is available in the spectrum, otherwise, we use the one based on the Mg\,II emission line.
For the remaining object (J0240$-$1851) that is not included in the SDSS catalog, the $M_{\rm BH}$ and $L_{\rm bol}$ are obtained from \cite{Muzahid2012} and \cite{Arav2013}, respectively.
The sample covers the ranges of $M_{\rm BH} \sim 10^{8.9} - 10^{10.4}~M_{\odot}$ and $L_{\rm bol} \sim 10^{46.3} - 10^{47.7}$~erg\,s$^{-1}$.
The $L/L_{\rm Edd}$, ranging from 0.05 to 1.5, is calculated based on these $L_{\rm bol}$ and $M_{\rm BH}$.
Both the $M_{\rm BH}$ and the $L/L_{\rm Edd}$ have a typical uncertainty of about $\pm$0.5\,dex \citep{Laor1998,Park2012,Shen2024}, as the uncertainty in $L/L_{\rm Edd}$ mainly comes from that in $M_{\rm BH}$, which is caused by the systematic uncertainties in the methods, and the uncertainty in $L_{\rm bol}$ is negligible compared to that in $M_{\rm BH}$ \citep{Richards2006}.

\section{Data Reduction}

The sample was observed with the VLA A configuration at 5.5 and 9.0\,GHz with a bandwidth of 2\,GHz and an angular resolution of 0.20\arcsec--0.33\arcsec.
At $z$ = 0.4--2.8, the corresponding spatial scales are $\sim$ 1--3\,kpc.
These observations were carried out between July and August 2023 with a total time of 7 hours for 13 objects in two bands, which yield an integrated time of approximately 10 minutes on each target in each band and an image sensitivity of about 10\,$\mu$Jy\,beam$^{-1}$.
The data were calibrated using the VLA calibration pipeline version 6.4 and the Common Astronomy Software Applications (CASA) version 6.5.
A standard flux density calibrator, either 3C\,48, 3C\,147, or 3C\,286, was used for each target.

To produce the radio maps, we used a cell size of 0.05\arcsec to properly sample the beam that has a FWHM of $\sim$ 0.3\arcsec.
The maps were created in a region of 1024 $\times$ 1024 pixels centered on the source coordinates to check for the presence of extended emission or nearby sources.
We modeled the main target, along with the nearby sources if present, using the CLEAN algorithm.
If a bright source is outside the mapped region, we enlarged the image to include the bright source to reduce the effect of its side-lobes.
We created the images using natural weighting, which maximizes the sensitivity at the expense of compromising the angular resolution.
Self-calibration was not applied to the targets due to the insufficient signal-to-noise (S/N) ratio.

We modeled the source with a Gaussian fit on the image plane and deconvolved it from the beam, to recover the radio position, the source size and position angle, and the integrated and peak flux densities, $F_{\rm total}$ and $F_{\rm peak}$, centered at 5.5 and 9.0\,GHz.
The background root mean square (RMS) was estimated in a source-free region.
If the source size is too small to be determined compared to the beam size, we adopted half of the beam size as an upper limit on the source size.
To measure the spectral slope with a minimal bias from resolution differences between the C and X bands, we tapered the images to the same $uv$ range, typically $\sim$ 20--660\,k$\lambda$ in both bands, corresponding to the angular coverage of $\sim$ 0.4\arcsec--13\arcsec.

The detection criterion we adopted is $> 3\sigma$ on the flux density, where $\sigma$ is the background noise level.
When the source is not detected, a 3$\sigma$ upper limit is reported.
In total, nine of the 13 objects in the sample are detected in both bands, with flux densities ranging from 0.02\,mJy to 0.8\,mJy.
The other four objects were not detected in either band. 
The VLA coordinates are consistent with the Gaia positions \citep{Gaia2016, Gaia2023} with a separation of $<$ 75\,mas, which is within the resolution of the VLA A configuration.
The radio maps of the nine detected objects at 5.5 and 9.0\,GHz are present in Figure~\ref{maps}.
The radio coordinates and source sizes are listed in Table~\ref{size}, and the flux densities in the full-array maps and the tapered maps are listed in Table~\ref{flux}.


\setcounter{figure}{0}
\begin{figure*}[ht!]
\centering
\includegraphics[width=1.35\columnwidth, trim={0cm, 1cm, 0cm, 1cm}, clip]{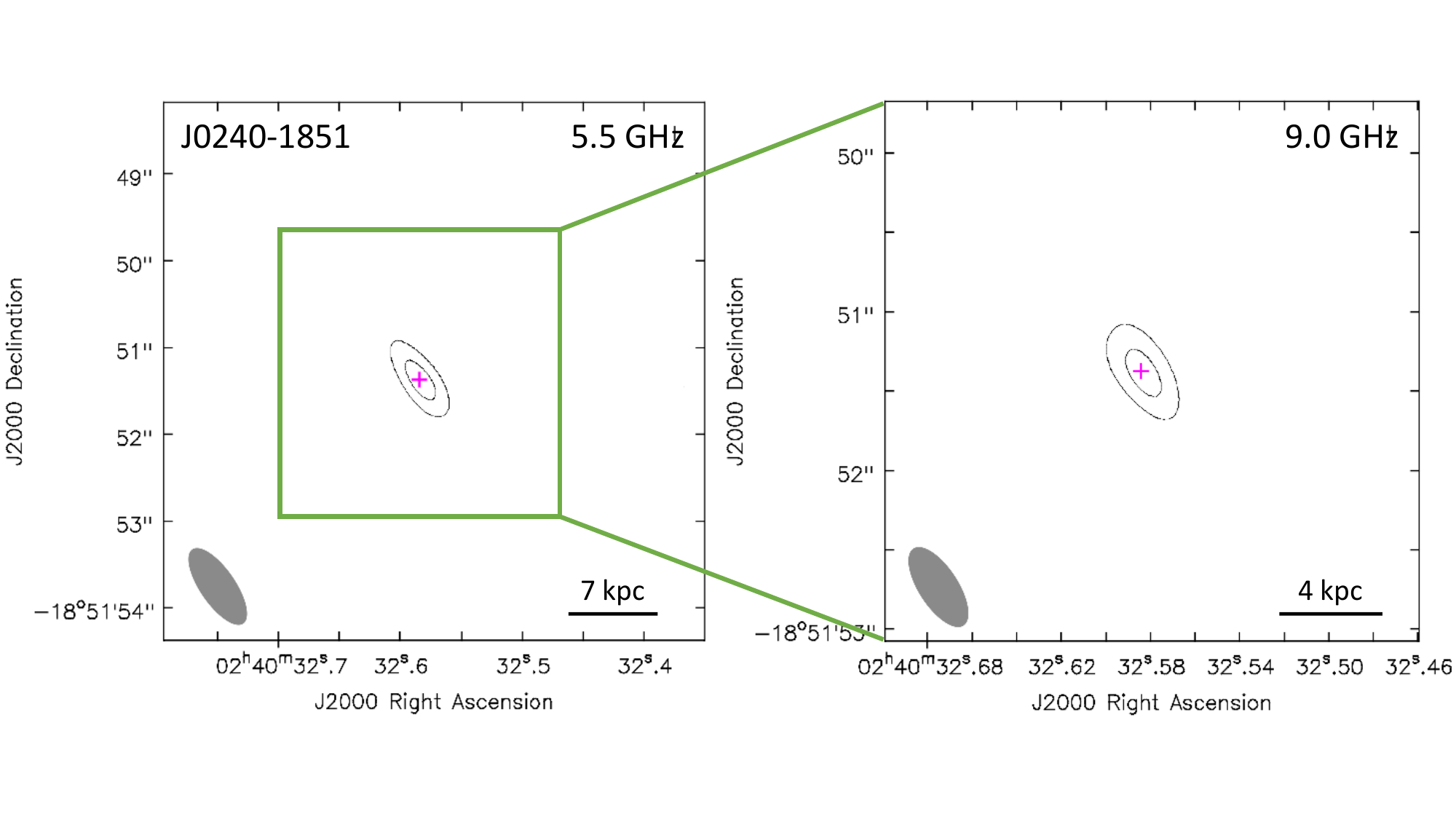}
\includegraphics[width=1.35\columnwidth, trim={0cm, 1cm, 0cm, 1cm}, clip]{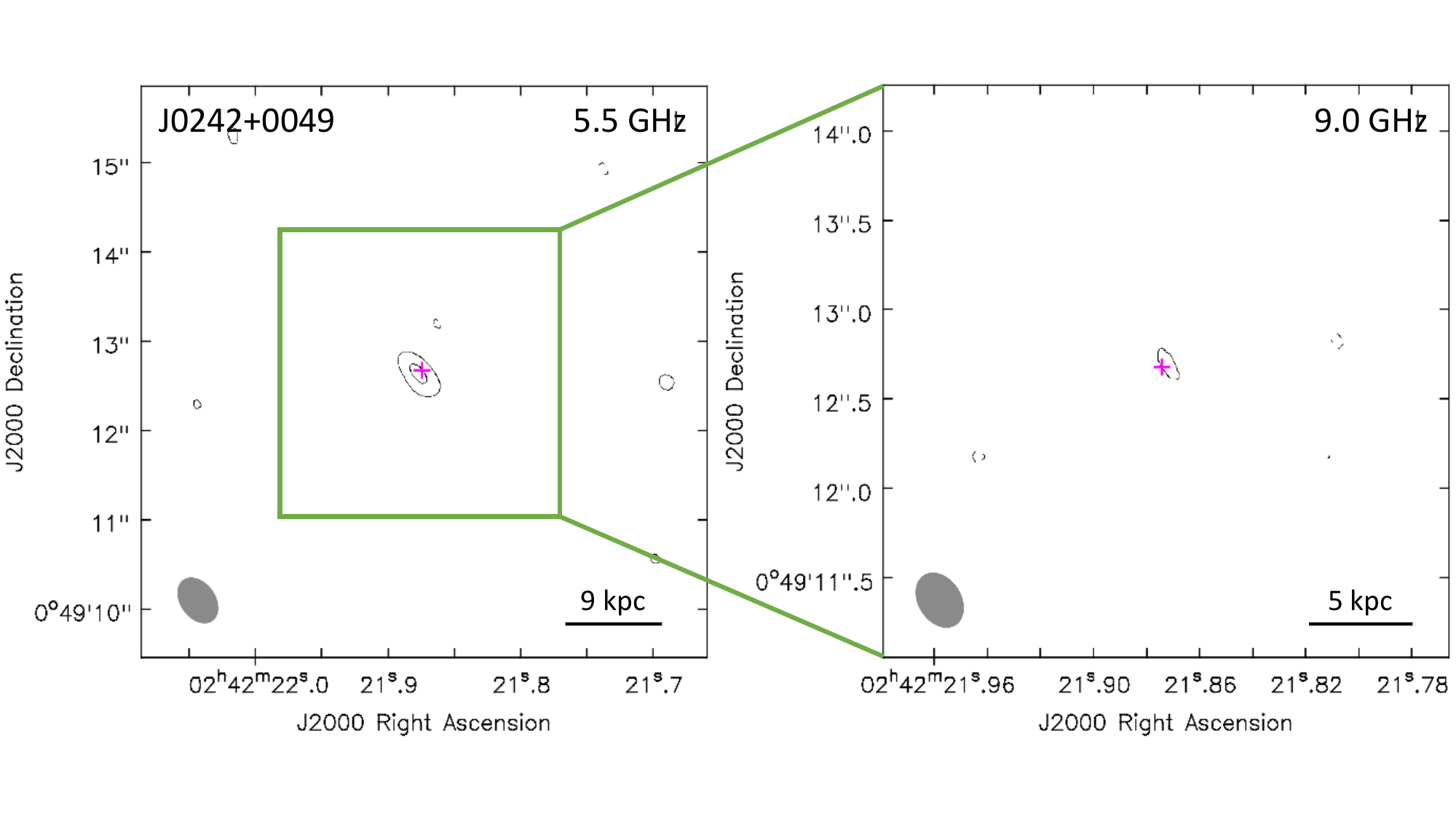}
\includegraphics[width=1.35\columnwidth, trim={0cm, 1cm, 0cm, 1cm}, clip]{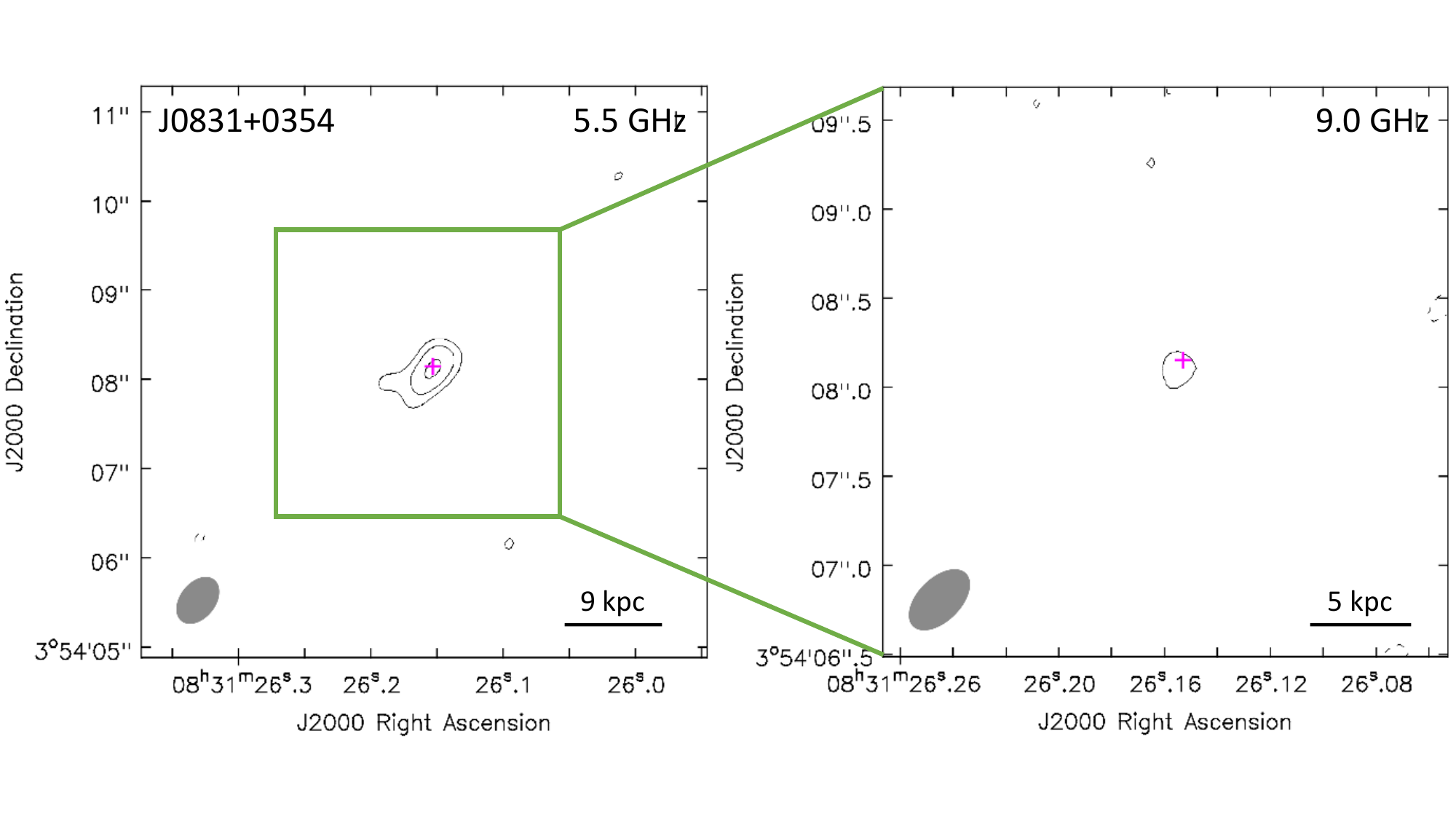}
\caption{The nine absorption-line quasars detected with the VLA A configuration at 5.5 and 9.0~GHz.
J0240$-$1851 (upper panel): The contours are at [$-$3, 3, 5] $\times$ 0.058~mJy\,beam$^{-1}$ at 5.5~GHz and [$-$3, 3, 5] $\times$ 0.066~mJy\,beam$^{-1}$ at 9.0~GHz.
J0242+0049 (middle panel): The contours are at [$-$3, 3, 5] $\times$ 0.008~mJy\,beam$^{-1}$ at 5.5~GHz and [$-$3, 3] $\times$ 0.007~mJy\,beam$^{-1}$ at 9.0~GHz.
J0831+0354 (lower panel): The contours are at [$-$3, 3, 5, 8] $\times$ 0.009~mJy\,beam$^{-1}$ at 5.5~GHz and [$-$3, 3] $\times$ 0.009~mJy\,beam$^{-1}$ at 9.0~GHz.}
\label{maps}
\end{figure*}

\setcounter{figure}{0}
\begin{figure*}[ht!]
\centering
\includegraphics[width=1.35\columnwidth, trim={0cm, 1cm, 0cm, 1cm}, clip]{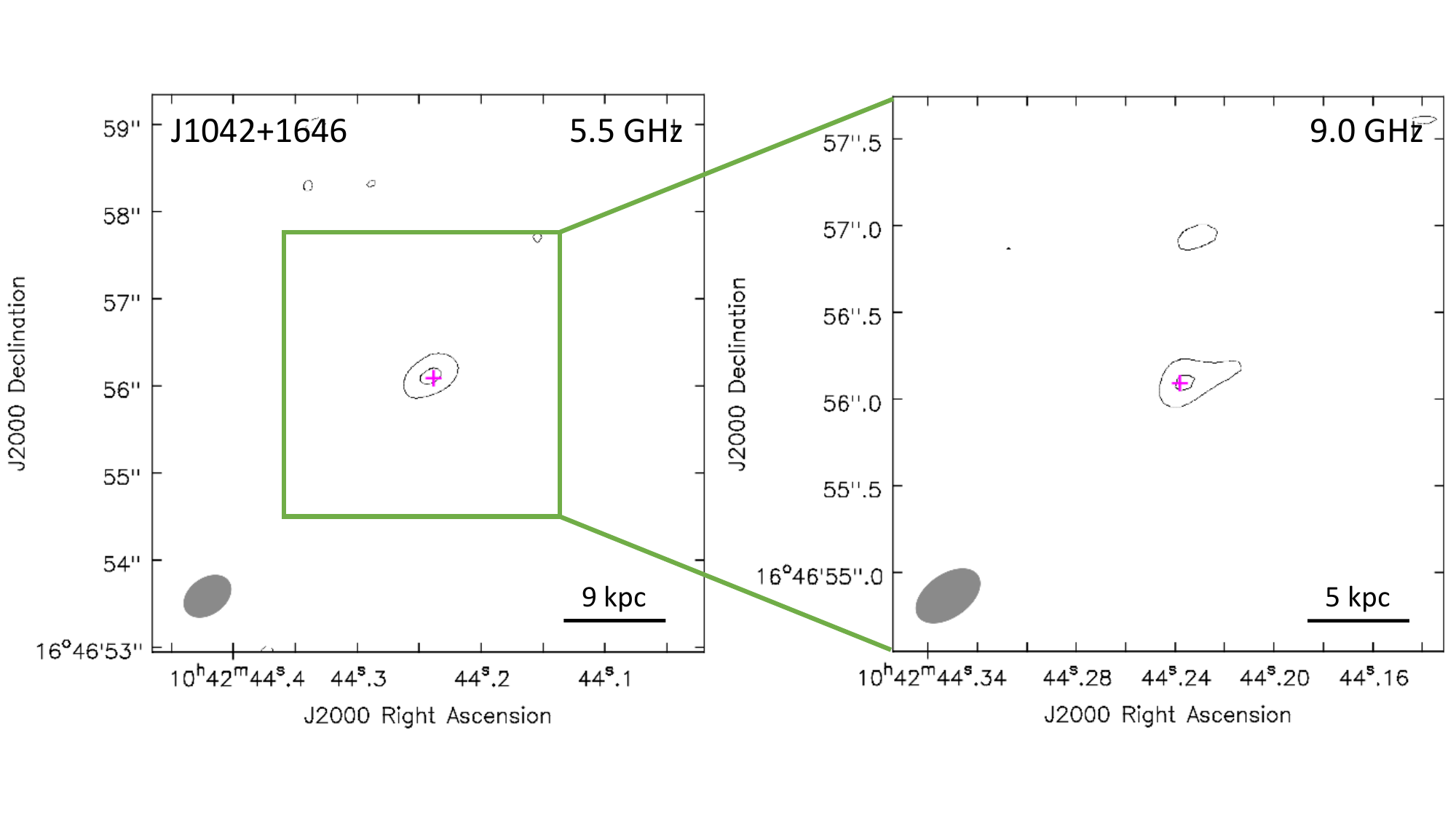}
\includegraphics[width=1.35\columnwidth, trim={0cm, 1cm, 0cm, 1cm}, clip]{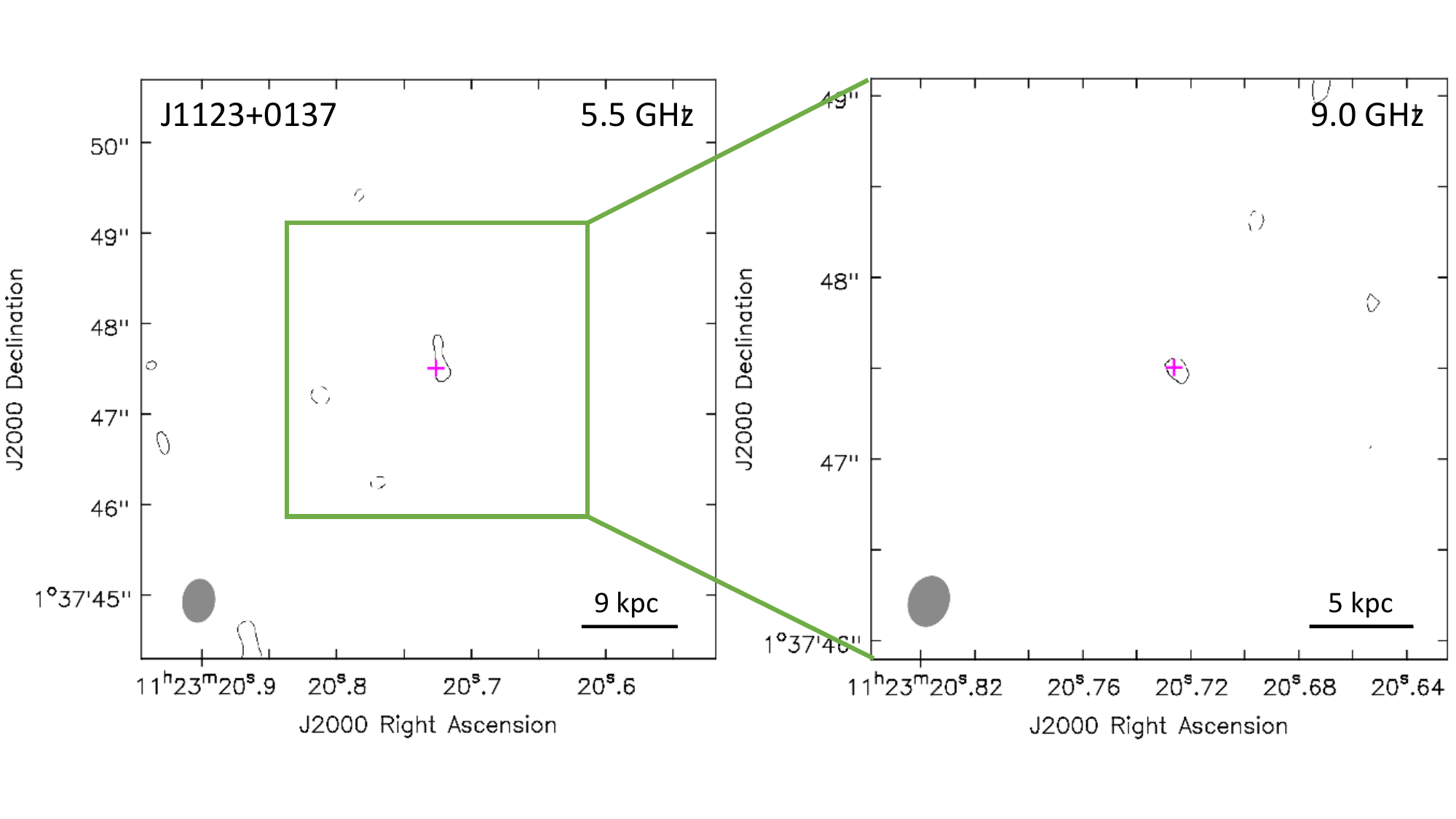}
\includegraphics[width=1.35\columnwidth, trim={0cm, 1cm, 0cm, 1cm}, clip]{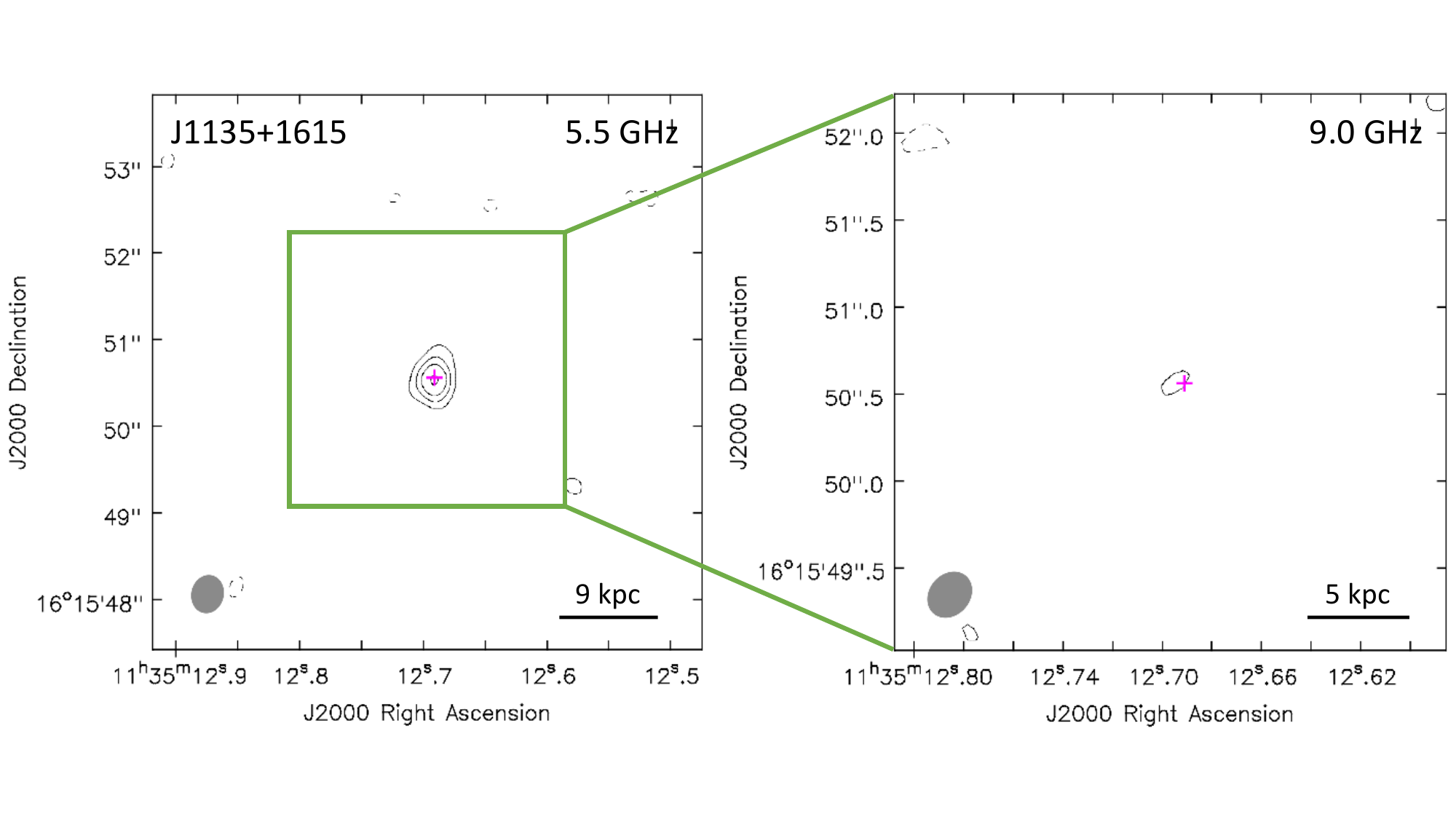}
\caption{Continued.
J1042+1646 (upper panel): The contours are at [$-$3, 3, 6] $\times$ 0.010~mJy\,beam$^{-1}$ at 5.5~GHz and [$-$3, 3, 5] $\times$ 0.010~mJy\,beam$^{-1}$ at 9.0~GHz.
J1123+0137 (middle panel): The contours are at [$-$3, 3] $\times$ 0.012~mJy\,beam$^{-1}$ at 5.5~GHz and [$-$3, 3] $\times$ 0.011~mJy\,beam$^{-1}$ at 9.0~GHz.
J1135+1615 (lower panel): The contours are at [$-$3, 3, 5, 7, 10] $\times$ 0.010~mJy\,beam$^{-1}$ at 5.5~GHz and [$-$3, 3] $\times$ 0.012~mJy\,beam$^{-1}$ at 9.0~GHz.}
\end{figure*}

\setcounter{figure}{0}
\begin{figure*}[ht!]
\centering
\includegraphics[width=1.35\columnwidth, trim={0cm, 1cm, 0cm, 1cm}, clip]{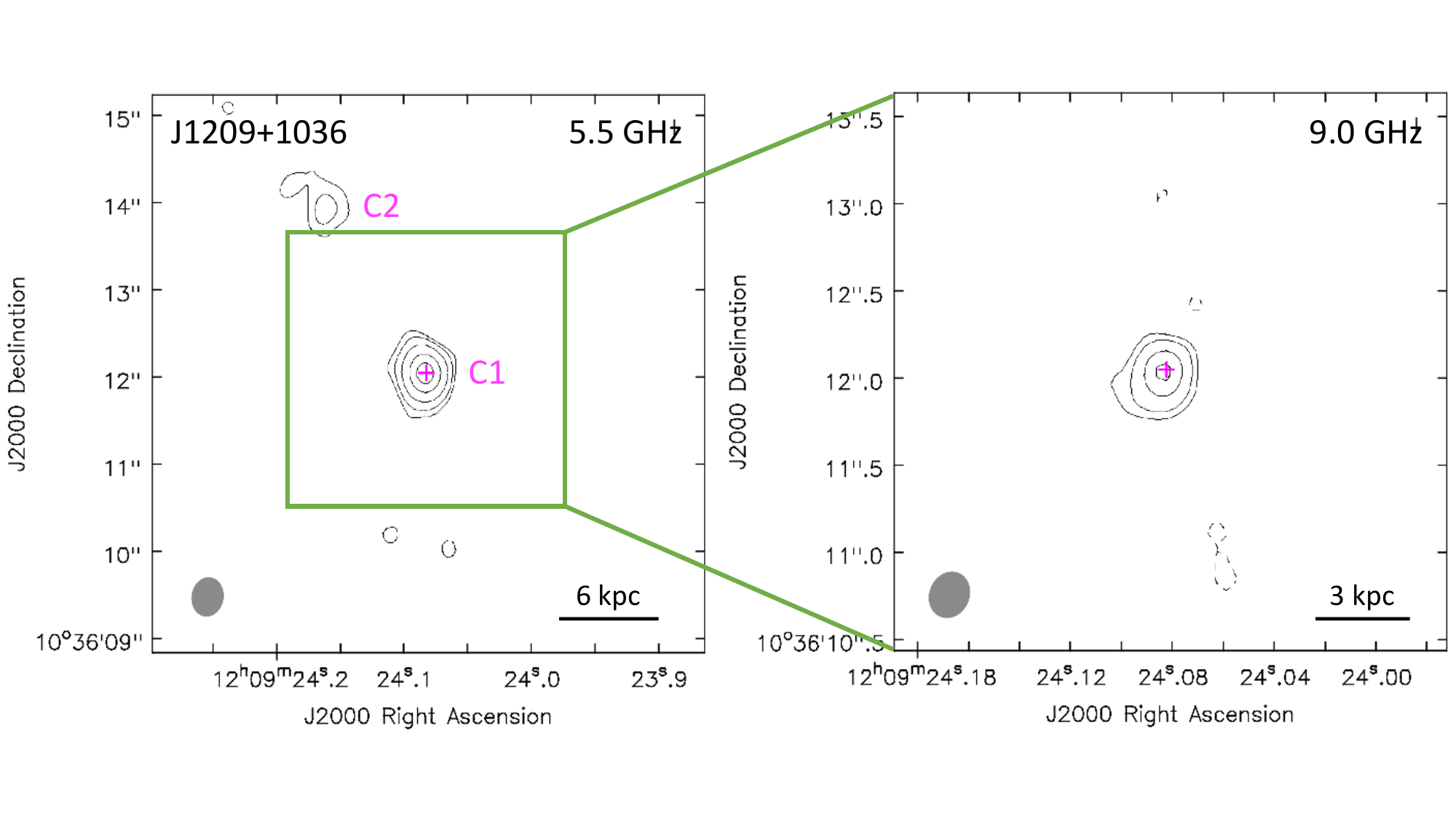}
\includegraphics[width=1.35\columnwidth, trim={0cm, 1cm, 0cm, 1cm}, clip]{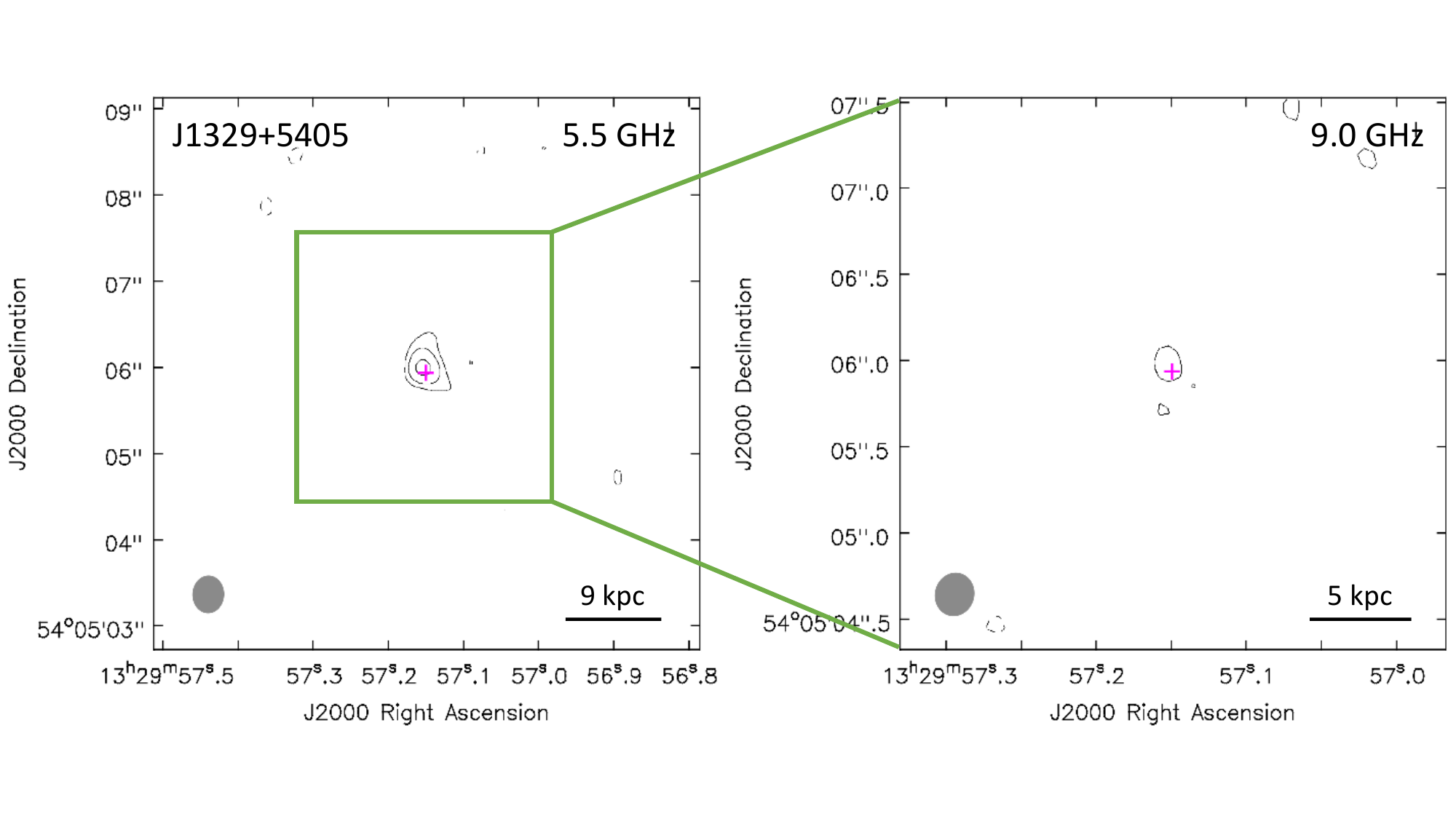}
\includegraphics[width=1.35\columnwidth, trim={0cm, 1cm, 0cm, 1cm}, clip]{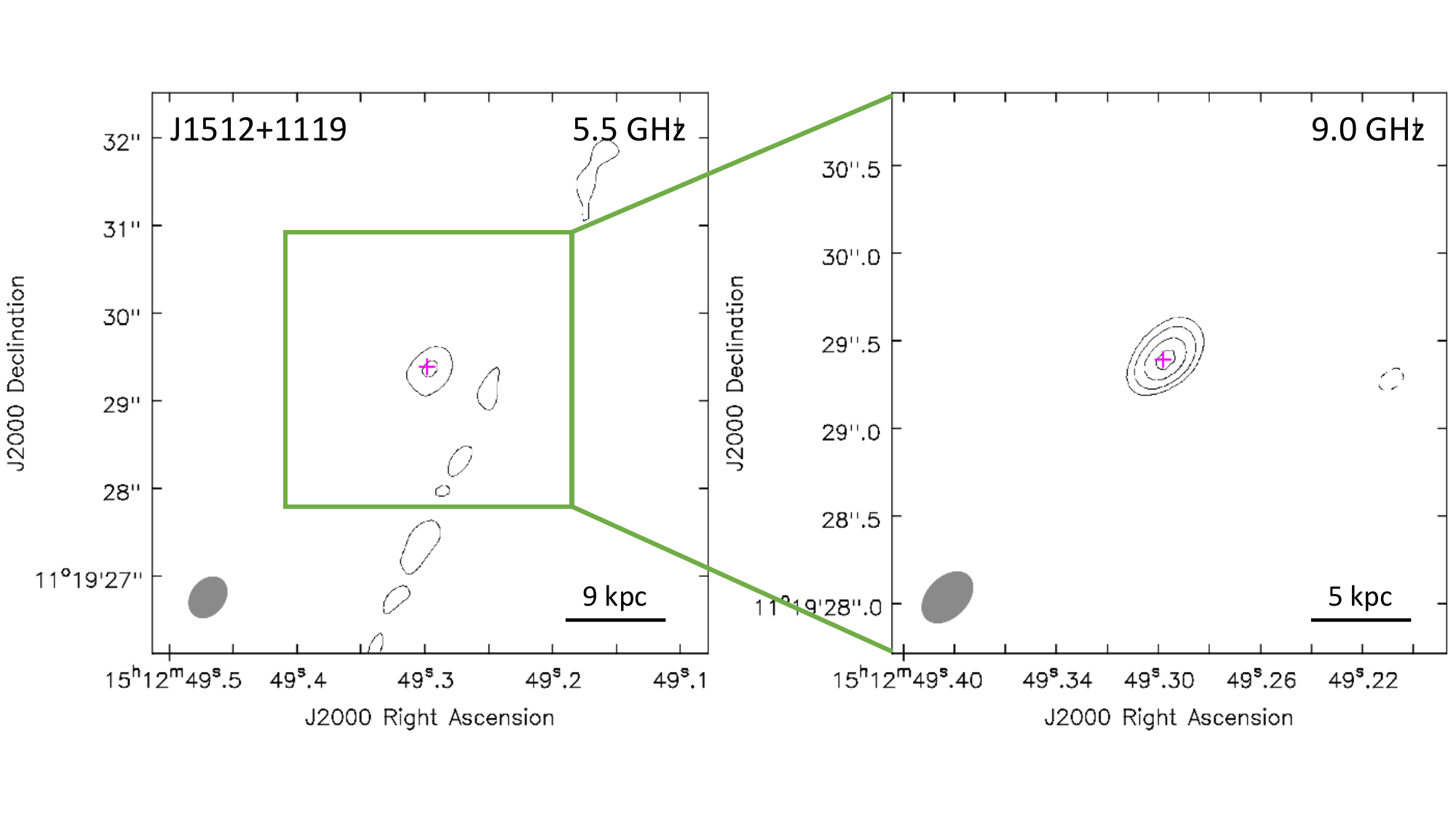}
\caption{Continued.
J1209+1036 (upper panel): The contours are at [$-$3, 3, 5, 10, 20, 30] $\times$ 0.011~mJy\,beam$^{-1}$ at 5.5~GHz and [$-$3, 3, 5, 10, 15] $\times$ 0.012~mJy\,beam$^{-1}$ at 9.0~GHz.
J1329+5405 (middle panel): The contours are at [$-$3, 3, 5, 8] $\times$ 0.011~mJy\,beam$^{-1}$ at 5.5~GHz and [$-$3, 3] $\times$ 0.012~mJy\,beam$^{-1}$ at 9.0~GHz.
J1512+1119 (lower panel): The contours are at [$-$3, 3, 6] $\times$ 0.014~mJy\,beam$^{-1}$ at 5.5~GHz and [$-$3, 3, 5, 8, 11] $\times$ 0.010~mJy\,beam$^{-1}$ at 9.0~GHz.}
\end{figure*}

\section{Data Analysis}

The VLA spectral slope at 5.5--9.0\,GHz, $\alpha_{5.5-9.0}$, of the nine detected objects is measured based on the $F_{\rm peak}$ in the tapered maps, which have comparable resolutions and cover emission on similar scales at both frequencies.
The slope cannot be determined for the four objects not detected at either frequency.
These observed 5.5--9.0\,GHz slopes correspond to the rest-frame slope of 8--13\,GHz at $z$ = 0.4 and of 17--28\,GHz at $z$ = 2.1.
The rest-frame 5.5 and 9.0\,GHz radio luminosities, $L_{\rm 5.5GHz}$ and $L_{\rm 9.0GHz}$, are also calculated based on the observed $\alpha_{5.5-9.0}$ and the $F_{\rm total}$ in the full-array maps.

The $\alpha_{5.5-9.0}$ at the rest frame may be somewhat steeper or flatter than that at the observed frame, as one can see in \cite{Baldi2022} which shows the radio spectrum of some RQ PG quasars up to 45~GHz.
The implied systematic uncertainty in $\alpha_{5.5-9.0}$ is $\Delta \alpha \sim \pm 0.5$, and the derived luminosity uncertainty is therefore within a factor of $\pm 2$.
Both uncertainties are too small to affect the trends discussed in Section 5.

The $R$ value is the rest-frame flux ratio of the radio at 5\,GHz to the optical in the B band.
The radio flux at the rest-frame 5\,GHz is derived based on the observed $\alpha_{5.5-9.0}$ and the $F_{\rm total}$ in the full-array maps.
If $\alpha_{5.5-9.0}$ is not available, we use a uniform slope of $\alpha = -0.5$.
The optical flux in the rest-frame B band is taken from the infrared photometric measurements in the NASA/IPAC Extragalactic Database (NED), mostly from the SDSS $i$, $r$, or $z$ bands, or the 2MASS J and H bands, depending on the redshift of the objects.

The $R$ value can be overestimated due to reddening, and BALQs indeed tend to be more reddened in the EUV than non-BALQs \citep[e.g.][]{Baskin2013}.
However, our objects are selected to be bright and luminous in the EUV, so are likely not significantly extincted.
The effect of the dust absorption in the optical is nearly ten times smaller than that in the EUV \citep{Laor1993}, and thus the reddening in the B band and its effect on $R$ are likely negligible.

All the objects are RQ with $R$ = 0.02--4.55.
The $\alpha_{5.5-9.0}$ of the nine detected objects have a wide span from $-$1.7 to 0.8.
The $L_{\rm 5.5GHz}$ and $L_{\rm 9.0GHz}$ are in the range of $10^{40.0}-10^{42.1}$~erg\,s$^{-1}$, which are about $10^{-6}\,L_{\rm bol}$.
The upper limits on the $R$ and the radio luminosity of the undetected objects also lie in the same ranges.
These values are reported in Table~\ref{result}.

We further compare our BALQ sample with the PG quasar sample \citep{Boroson1992}, which is the most extensively studied Type 1 AGN sample, of which the radio slope $\alpha_{5.5-9.0}$ in 25 $z < 0.5$ RQ PG quasars was studied systematically \citep{Laor2019}.
The study found a highly significant correlation of $\alpha_{5.5-9.0}$ with $L/L_{\rm Edd}$, which provides important hints on the radio emission mechanisms.
The ranges of redshift and $M_{\rm BH}$ of these two samples distribute differently, while the $L/L_{\rm Edd}$ has a similar distribution.
A comparison on the $\alpha_{5.5-9.0}$ can be made between the two samples, as the radio slope in RQ quasars reflects the radio emission mechanisms, which is likely independent of the redshift and not correlated with the $M_{\rm BH}$ \citep[][Figure 1 there]{Laor2019}.

\section{Results}

The radio emission in all the nine detected objects is mostly compact and is constrained within a scale of $<$ 1--4\,kpc, with $F_{\rm peak}/F_{\rm total} \gtrsim 0.7$ at 5.5\,GHz and $F_{\rm peak}/F_{\rm total} \gtrsim 0.5$ at 9.0\,GHz.
The higher $F_{\rm peak}/F_{\rm total}$ at 5.5\,GHz than at 9.0\,GHz is likely due to the lower resolution at 5.5\,GHz than at 9.0\,GHz.
The fact that the detected radio emission is generally compact can also be seen from a flux comparison between the full-array and the tapered maps, which show comparable fluxes with different $uv$ coverages.

\begin{figure}[ht!]
\centering
\includegraphics[width=\columnwidth, trim={0cm, 0cm, 1cm, 1cm}, clip]{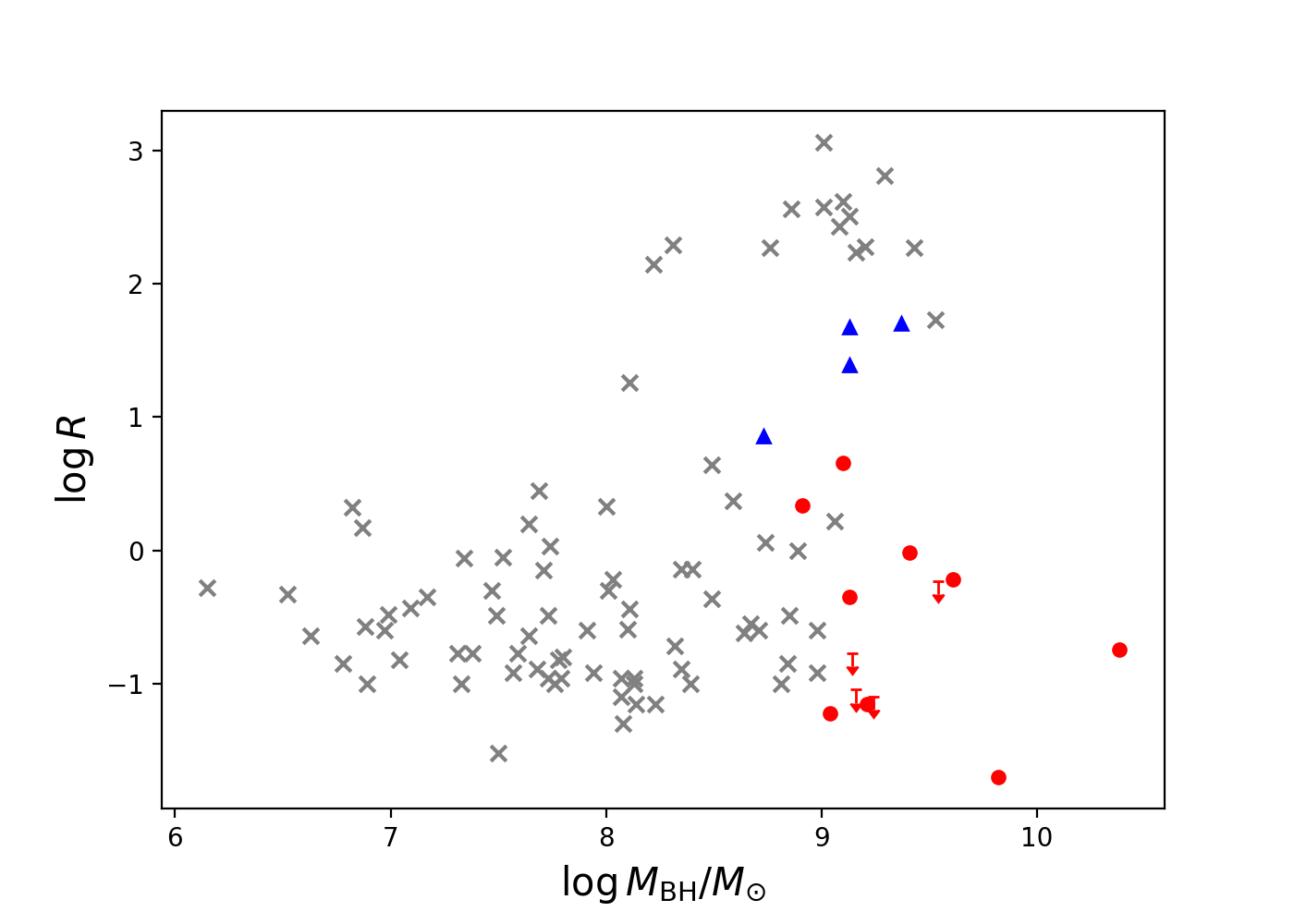}
\caption{The radio loudness versus the BH mass. The red circles and down-arrows represent the BALQs observed in our VLA observations. The blue triangles represent the BALQs excluded in our VLA observations. The grey crosses represent the whole PG quasar sample. The $M_{\rm BH}$ uncertainty is typically $\pm$0.5\,dex.
The $R$ values of the RQ BALQs are comparable to those of the RQ PG quasars, suggesting that the winds in BALQs do not produce stronger radio emission than those in non-BALQs.}
\label{RL}
\end{figure}

\begin{figure}[ht!]
\centering
\includegraphics[width=\columnwidth, trim={0cm, 0cm, 1cm, 1cm}, clip]{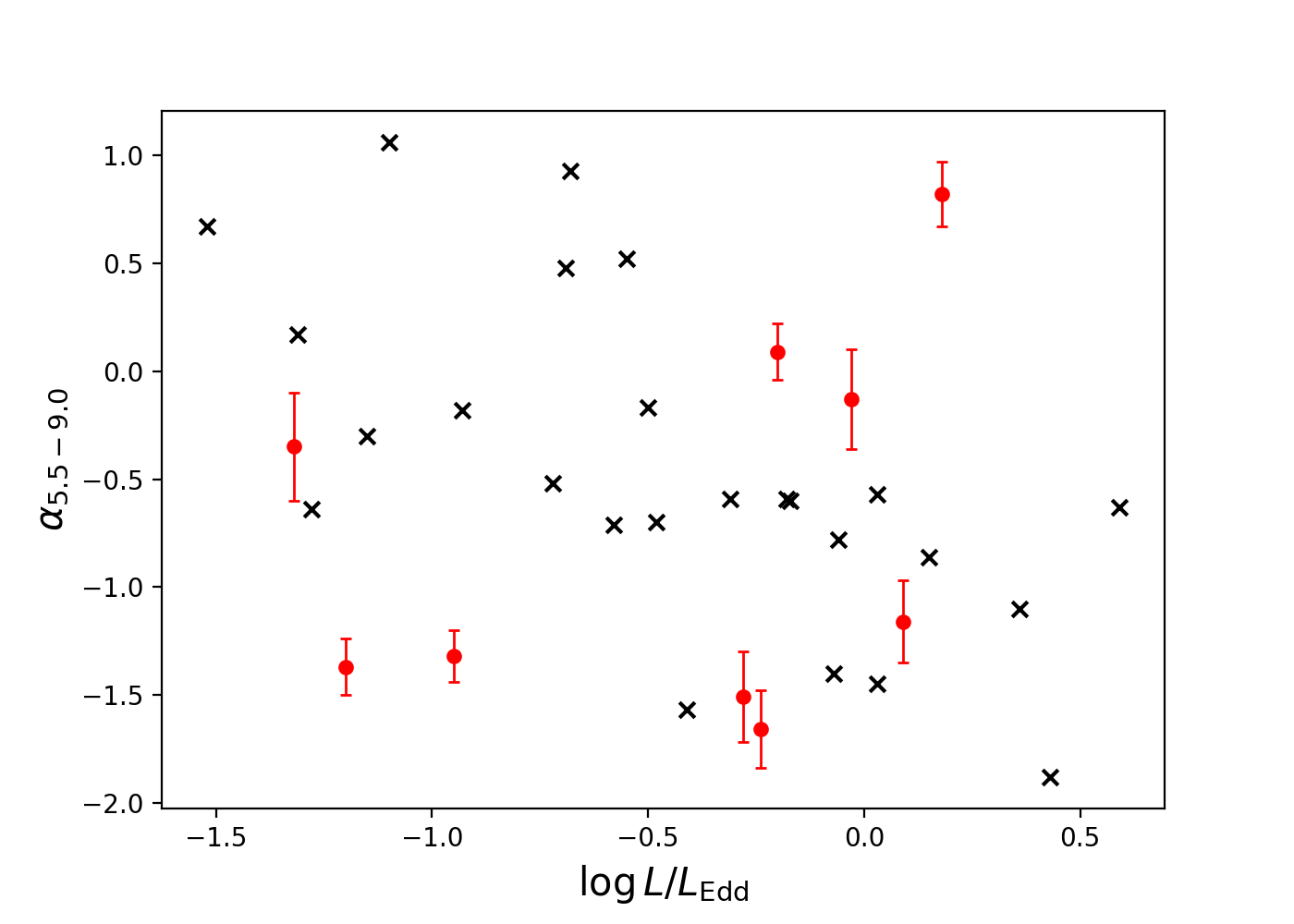}
\caption{The observed 5.5--9.0~GHz slope versus the Eddington ratio. The red circles represent the RQ BALQs detected in our VLA observations. The black crosses represent the 25 RQ PG quasars which the $\alpha_{5.5-9.0}$ was studied systematically \citep{Laor2019}. The $L/L_{\rm Edd}$ uncertainty is typically $\pm$0.5\,dex.
The RQ BALQs generally does not follow the $\alpha_{5.5-9.0}$ and $L/L_{\rm Edd}$ correlation found in the RQ PG quasars, where high and lower $L/L_{\rm Edd}$ objects exhibit steep and flat slopes, respectively.}
\label{slope}
\end{figure}

Figure~\ref{RL} shows the radio loudness versus the BH mass of our BALQ sample (including 13 objects with $R$ = 0.02--4.55 observed in our VLA observations and 4 objects with $R$ = 7--51 not observed) and the whole PG quasar sample.
The $R$ values of the RQ BALQs are comparable to that of the RQ PG quasars ($R$ = 0.03--4.37).
However, the $M_{\rm BH}$ distributions of the two samples are distinct, with $\log M_{\rm BH}/M_{\odot}$ = 6.2--9.5 for the PG quasars and $\log M_{\rm BH}/M_{\odot}$ = 8.7--10.4 for our BALQs.
The high $M_{\rm BH}$ in the BALQs is a selection effect, as they are selected to be bright at high redshift, and thus luminous.
The RL fraction of the BALQs differs significantly from that of the PG quasars, as at $M_{\rm BH} > 10^9 M_{\odot}$, the RL fraction is 91\% (10/11) \citep{Laor2000} for the PG quasars, while it is 20\% (3/15) for the BALQs.

Figure~\ref{slope} shows the observed 5.5--9.0\,GHz slope versus the Eddington ratio of the nine detected BALQs and the 25 RQ PG quasars where the $\alpha_{5.5-9.0}$ was studied systematically \citep{Laor2019}.
The distribution of the radio spectral slopes of the two samples are comparable ($\alpha_{5.5-9.0}$ from $-$1.7 to 0.8 for the RQ BALQs and from $-$1.9 to 1.1 for the RQ PG quasars), which is in agreement with \cite{Barvainis1997} who also found a similar distribution in the radio spectral shapes at 1.5--14.9\,GHz of RQ BALQs and non-BALQs.
In these nine objects, six have high $L/L_{\rm Edd}$ ($\gtrsim$ 0.5) and three have lower $L/L_{\rm Edd}$ ($\sim$ 0.1).
The RQ BALQs generally does not follow the $L/L_{\rm Edd}$ and radio spectral slope correlation found in the RQ PG quasars \citep{Laor2019}, where high and lower $L/L_{\rm Edd}$ objects have steep and flat slopes, respectively.

\begin{figure}[ht!]
\centering
\includegraphics[width=\columnwidth, trim={4cm, 0cm, 4cm, 0cm}, clip]{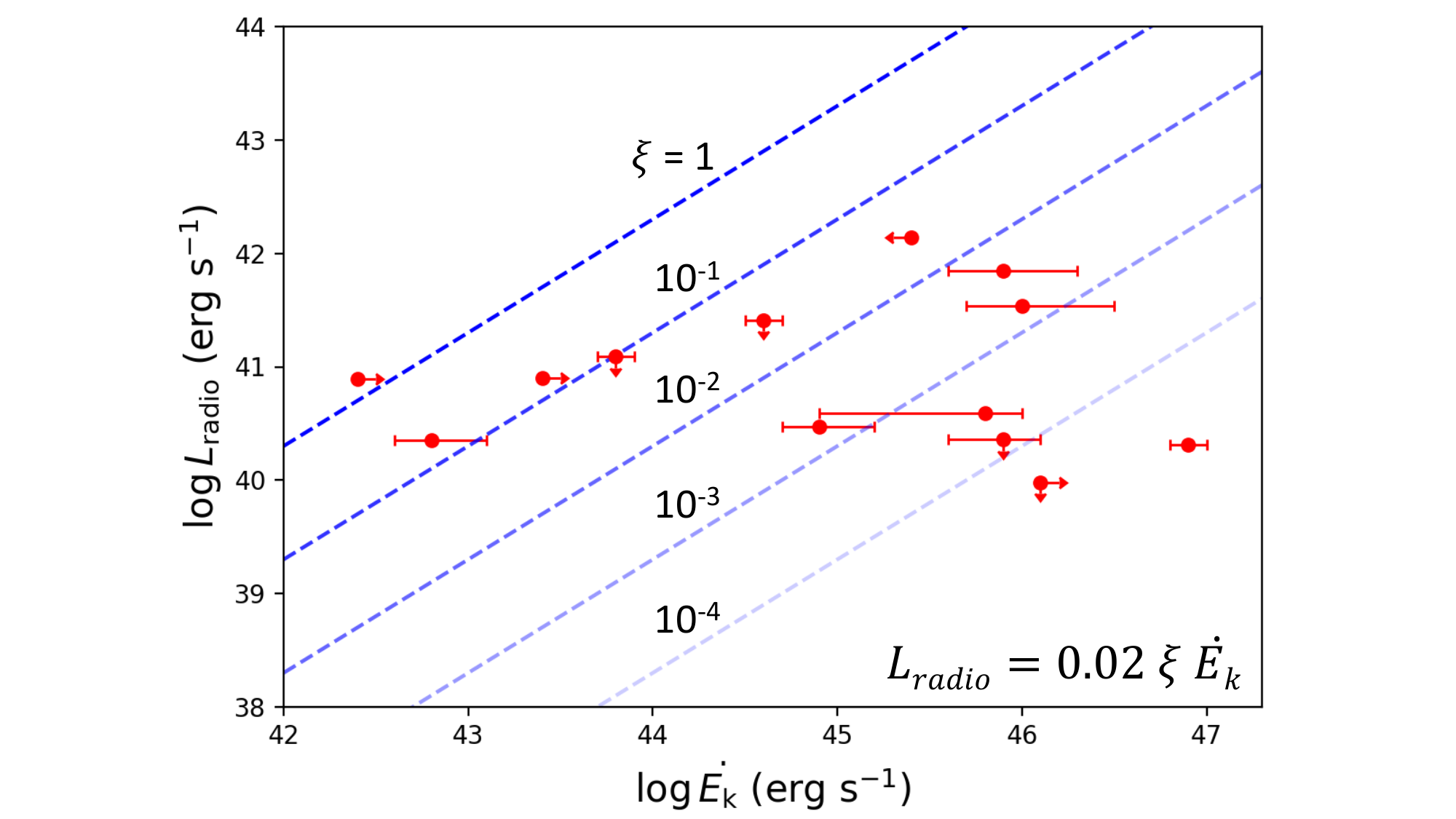}
\caption{The radio 5.5\,GHz luminosity versus the kinetic power.
The blue dashed lines represent the predicted synchrotron emission produced by an AGN outflow based on the expression in \citet[][Eq.32 therein]{Nims2015} as shown in the right-bottom corner.
The value of $\xi$, which is marked next to each line, represents the fraction of the outflow kinetic power converted into the relativistic electron power that is proportional to the radio synchrotron luminosity.
The lack of a correlation between the radio luminosity and the kinetic power suggests that the detected radio emission is not entirely produced by the wind shock.}
\label{luminosity}
\end{figure}

Figure~\ref{luminosity} shows the radio 5.5\,GHz luminosity versus the kinetic power of our BALQ sample.
We further compare the observed radio emission with the predicted synchrotron emission produced by an AGN wind shock model in \citet[][Equation (32) there]{Nims2015}.
The model assumes an energy-conserving wind, which travels at 10\% the speed of light and can reach $>$ 1\,kpc scales in a typical ambient medium, and a conversion rate $\xi$, which represents the fraction of the outflow $\dot{E}_{\rm k}$ converted into the relativistic electron power that is proportional to the radio synchrotron luminosity.
At supernova shocks, $\xi \approx 10^{-2}$ typically, that is about 1\% of the shock energy going into relativistic electrons \citep{Thompson2006}.
There is no obvious difference in the physical mechanisms of AGN-driven shocks compared to supernova shocks that would alter the average electron acceleration efficiency.

We find that the radio luminosity does not correlate with the $\dot{E}_{\rm k}$ as predicted by the model.
The radio luminosity has a narrower range of $10^{40.0}-10^{42.1}$~erg\,s$^{-1}$, while the $\dot{E}_{\rm k}$ spans a wider range of $10^{42.4}-10^{46.9}$~erg\,s$^{-1}$, about 5 orders of magnitude.
The observed radio luminosity ($\sim 10^{-6}\,L_{\rm bol}$) is in an expected range for RQ quasars with $L_{\rm bol} = 10^{46}-10^{48}$~erg\,s$^{-1}$.
Our sample is optically selected, and thus is unbiased in terms of their radio properties (nine of the 13 objects were detected, and the upper limits for the non-detections are similar to the detected radio luminosities).
If the radio emission is produced by the AGN-driven shocks, the wide range of the $\dot{E}_{\rm k}$ results in the efficiency of the wind shock power converted into the relativistic electron power spanning about 5 orders of magnitude (from $10^{-5}$ to 1).
This suggests that the detected radio emission is not entirely produced by the wind shock, and that other mechanisms also contribute, provided there is no intrinsic difference between the radio emission produced by AGN-driven and supernova shocks.

\section{Discussion}

Where does the radio emission come from?
The sample was selected to show AGN-driven winds extending from $\sim$ 0.1\,pc up to $\sim$ 70\,kpc based on the spectroscopic diagnostics of the blueshifted UV absorption lines.
Do these outflows produce the observed radio emission through an interaction with the surrounding gas in the host galaxy?

The radio spectral slope $\alpha_{5.5-9.0}$ is steep ($< -0.5$) in five of the nine detected objects (Figure~\ref{slope}), indicating optically thin synchrotron emission which may come from an outflow.
Three objects, J0831+0354, J1135+1615, and J1329+5405, have a high $L/L_{\rm Edd}$ ($\gtrsim$ 0.5), suggesting a radiation-pressure-driven wind \citep{Laor2019}.
The other two objects, J0242+0049 and J1209+1036, have a lower $L/L_{\rm Edd}$ ($\lesssim$ 0.1), which may suggest that the wind is not radiation pressure driven (e.g., magnetically driven) or the radio emission is not associated with a wind (e.g., a low-power jet or star formation).

Interestingly, J1209+1036 is the only object exhibiting an extended component C2 with a separation of $\sim$ 12\,kpc from the core component C1.
This distance is a factor of $\sim$ 20 larger than the outflow size of $\sim$ 0.5\,kpc predicted by the UV spectroscopic measurement (Table~\ref{sample}).
The radio emission is thus not likely to be produced by the same outflow.
The object is the only RQ BALQ in our sample detected by the FIRST with $F_{\rm peak}$ = 1.91~mJy and $F_{\rm total}$ = 2.35~mJy at 1.4\,GHz.
The resolution of the FIRST (5\arcsec) is about 10 times lower than that of the VLA A configuration (0.3\arcsec).
The spectral slope at 1.4--5.5\,GHz is flatter than $-0.80$, which is calculated using the $F_{\rm peak}$ at 1.4\,GHz and $F_{\rm total}$ of the two components at 5.5\,GHz.
The 1.4--5.5\,GHz spectral slope and the 5.5\,GHz to 1.4\,GHz flux ratio ($\sim$ 0.3) indicate the presence of extended emission dominated on larger scales ($\sim$ a few kpc) and at lower frequencies ($\sim$ 1\,GHz).

The radio spectral slope $\alpha_{5.5-9.0}$ is flat ($> -0.5$) or inverted ($> 0$) in the other four of the nine detected objects (Figure~\ref{slope}).
If the emission is intrinsically optically thick, we can estimate the physical size of the radio synchrotron source via \citep[][Equation (22) there]{Laor2008}
\begin{equation}
R_{\rm RS} = 0.47 L_{30}^{0.4} L_{46}^{0.1} \nu_{\rm peak}^{-1}
\end{equation}
where $R_{\rm RS}$ is the radius of the radio source in pc, $\nu_{\rm peak}$ is the turnover frequency in GHz, $L_{30}$ is the radio luminosity density at the turnover frequency in 10$^{30}$\,erg\,s$^{-1}$\,Hz$^{-1}$, and $L_{46}$ is the bolometric luminosity in 10$^{46}$\,erg\,s$^{-1}$.
The derived radio source sizes of $\sim$ 0.1--0.2\,pc imply that the rest-frame 9\,GHz emission comes from a compact region on the accretion disk and the inner BLR scales, possibly originating from the accretion disk corona.

Alternatively, the emission may be intrinsically optically thin, but free–free absorbed by AGN photoionized gas.
The free-free optical depth scales as $\tau_{\rm ff} \propto \nu^{-2}$.
For example, if $\tau_{\rm ff}$\,(7\,GHz) = 1, then $\tau_{\rm ff}$\,(5.5\,GHz) = 1.62 and $\tau_{\rm ff}$\,(9.0\,GHz) = 0.60.
This differential absorption leads to a significant spectral flattening by
\begin{equation}
\Delta \alpha = - \frac{\log [e^{\tau_{\rm ff} \rm{(9.0\,GHz)}} / e^{\tau_{\rm ff} \rm{(5.5\,GHz)}}]}{\log (9.0/5.5)} = 2.1.
\end{equation}

The distance of the absorber from the center, $r$, can be estimated via \citep[][Equations (11) and (21) there]{Baskin2021}
\begin{equation}
\nu_{\rm thick} = 6.03 \times 10^{11} (\frac{r}{r_{\rm dust}})^{-0.95} \, \rm{Hz}
\end{equation}
for dusty gas, where $\nu_{\rm thick}$ is the rest-frame frequency at which $\tau_{\rm ff} = 1$ and assumed to be 7(1+$z$)\,GHz, and the dust sublimation radius is $r_{\rm dust} = 0.2 L_{46}^{0.5}$~pc.
The derived distance of the absorbing medium from the central source is of $\sim$ 37--62\,pc.

The assumption of $\tau_{\rm ff} = 1$ at the observed 7\,GHz is required, as the spectral flattening due to free-free absorption would not occur if $\tau_{\rm ff} = 1$ at frequencies below 5.5\,GHz or above 9.0\,GHz.
In the former case, the radio emission at 5.5 and 9.0~GHz would remain effectively unabsorbed, resulting in an intrinsically flat $\alpha_{5.5-9.0}$.
In the later case, the emission at these frequencies would be heavily absorbed, leading to a strongly inverted $\alpha_{5.5-9.0}$, or probably even non-detection given our detection limits.

In three of the four flat-spectrum objects, J1042+1646, J1123+0137, and J1512+1119, which reside at high $L/L_{\rm Edd}$ ($\gtrsim$ 0.6), the location of the free-free absorber is comparable to the outflow sizes of the low-velocity component, of about a few tens pc, as derived from the UV spectroscopy (Table~\ref{sample}).
Thus the UV wind could be extended enough to cover the radio source, and its free-free absorption can flatten the observed $\alpha_{5.5-9.0}$.
The other object, J0240$-$1851, resides at lower $L/L_{\rm Edd}$ ($\lesssim$ 0.1), and the required location of the free-free absorber is a factor of $\sim$ 40 smaller than the estimated outflow sizes, of about a few thousands pc, as derived from the UV spectroscopy (Table~\ref{sample}).
Therefore, the radio emission cannot be intrinsically optically thin, and is not likely associated with a radiation-driven wind.
It is more likely from a compact source, possibly originating from the accretion disk corona.

Our RQ BALQs have $R < 5$ (Figure~\ref{RL}), which is comparable to the RQ PG quasars, suggesting that the winds in the RQ BALQs do not produce stronger radio emission than those in the typical RQ quasars.
This is consistent with \cite{Petley2024} who found the same distribution of radio emission properties (e.g., the radio detection rate, radio loudness, and radio luminosity) in the BALQs and non-BALQs with the same reddening.
The absence of a direct link between the radio emission properties and the presence of an outflow can also be seen through the radio luminosity which does not increase with the increasing $\dot{E}_{\rm k}$ (Figure~\ref{luminosity}).

Essentially, most (15/17) of our BALQs have $M_{\rm BH} > 10^9\,M_{\odot}$, but most (12/15 = 80\%) of them are RQ.
This is in contrast to the PG quasars, where only a small fraction (1/11 = 9\%) of objects with such a massive BH are RQ \citep{Laor2000} --- a trend also seen in various other studies \citep{Dunlop2003,Best2005}.
Why do most of the BALQs with a massive BH are not RL?

One possibility is that the outflows may disrupt the formation or collimation of radio jets.
An anti-correlation between the equivalent width of the \ion{C}{4} absorption line and the optical to X-ray spectral slope $\alpha_{\rm ox}$ shows that the objects with higher wind velocities have steeper $\alpha_{\rm ox}$, that is weaker EUV and X-ray emission \citep{Brandt2000,Laor2002,Gibson2009,Timlin2021}, which is not expected in the thin accretion disk models.
A possible mechanism which suppresses the jet formation in strong wind objects, is that the wind drives away the hot coronal gas reducing the X-ray emission, and thus disrupts the jet formation, as suggested in various studies of the jet-wind connection \citep[e.g.][]{Jackson2025}.
If this is true, it may suggest that the coronal emission, both the radio and the X-ray, in the BALQs is weak.
In this case, the flat-spectrum radio emission is probably not due to the compact coronal synchrotron self-absorption, but rather an extended steep-spectrum source, such as a wind, which is flattened by the foreground free-free absorption gas.

Star formation produces both synchrotron emission from supernova remnants and free-free emission form \ion{H}{2} regions.
The resolution of the observations corresponds to the spatial scales of $\sim$ 1--3\,kpc, which may include a fraction of radio emission associated with the host SF activity.
However, the SF-related radio emission tends to dominate in the MHz regime and show a rather uniform steep slope with $\alpha \simeq -0.8$ in the GHz regime \citep{Magnelli2015,CalistroRivera2017,Gim2019,An2021}, which is not consistent with the slopes observed here.
Thus the contribution of SF is probably not significant at the observed frequencies.

\section{Summary}

We present the results of our VLA observations at 5.5\,GHz and 9.0\,GHz of a sample of 13 RQ BALQs, where the outflow $\dot{E}_{\rm k}$ is measurable through the UV spectroscopy.
Nine of them are detected.
The results are summarized as follows.

1. The radio emission is generally constrained within a scale of $<$ 1--4\,kpc.
This suggests that no significant extended emission comes from a galactic-scale wind beyond a few kpc.

2. In the nine detected objects, the radio spectral slope $\alpha_{5.5-9.0}$ is steep ($< -0.5$) in five objects and is flat or inverted ($> -0.5$) in four objects.
We present evidence that the steep-slope emission can be associated with the UV outflows, and that the flat-slope emission can be intrinsically steep but flattened by free-free absorption from the UV outflowing gas.

3. However, we find no correlation between the radio luminosity and the estimated outflow $\dot{E}_{\rm k}$, which suggests that the outflows are not a major source of the observed radio emission.
In addition, the $R$ value of these RQ BALQs is comparable to that of typical RQ quasars, implying that the UV outflows in BALQs likely do not produce stronger radio emission compared to those in non-BALQs.

Future higher resolution radio observations on $\sim$ 100-pc scales may spatially resolve the radio emission if it originates from a wind.
Future multi-frequency radio observations can test the free-free absorption interpretation, as the emission should be completely absorbed below 5\,GHz where $\tau_{\rm ff} \gg 1$, and become optically thin above 9\,GHz where $\tau_{\rm ff} \ll 1$, as found in some nearby AGN \citep{Baskin2021}.
If this is verified, radio observations on $\sim$ 100-mas scales can be used to map the absorbing wind column density distribution.

\begin{acknowledgments}

We thank the anonymous referee for suggestions leading to the improvement of this work.
S.C. is supported in part at the Technion by a fellowship from the Lady Davis Foundation.
A.L. acknowledges support by the Israel Science Foundation (grant no.1008/18).
The Technion team is supported by a grant from the U.S.-Israel Binational Science Foundation (BSF) and the U.S. National Science Foundation (NSF).
The National Radio Astronomy Observatory is a facility of the National Science Foundation operated under cooperative agreement by Associated Universities, Inc.

\end{acknowledgments}

\vspace{5mm}
\facilities{VLA}
\software{CASA \citep{CASA2022}, Astropy \citep{Astropy2022}}

\bibliography{main.bbl}
\bibliographystyle{aasjournal}


\begin{turnpage}
\begin{table}[ht!]
\caption{The redshift, bolometric luminosity, BH mass, Eddington ratio, kinetic power, and outflow sizes of the sample.}
\label{sample}
\centering
\footnotesize
\begin{tabular}{llccccccll}
\hline\hline
Name & Full name & $z$ & Scale & Type & $\log L_{\rm bol}$ & $\log \frac{M_{\rm BH}}{M_{\odot}}$ & $\frac{L}{L_{\rm Edd}}$ & $\log \dot{E}_{\rm k}$ & $r_{\rm out}$ \\
& & & (kpc\,arcsec$^{-1}$) & & (erg\,s$^{-1}$) & & & (erg\,s$^{-1}$) & (pc) \\
(1) & (2) & (3) & (4) & (5) & (6) & (7) & (8) & (9) & (10) \\
\hline
J0240$-$1851 & 2MASS J02403258$-$1851512 & 0.631 & 7.03 & mini-BALQ & 47.18 & 10.38 & 0.05 & $45.8_{-0.9}^{+0.2}$ & $2000_{-320}^{+1200}, 3400_{-490}^{+2000}$ \\
J0242+0049 & SDSS J024221.87+004912.6 & 2.057 & 8.56 & mini-BALQ & 46.78 & 9.61 & 0.11 & $46.0_{-0.3}^{+0.5}$ & $<5400^{+7300}, 1200_{-900}^{+800}, 67000_{-31000}^{+55000}$ \\
J0831+0354 & SDSS J083126.15+035408.0 & 2.075 & 8.55 & BALQ & 47.11 & 8.91 & 1.23 & $45.9_{-0.3}^{+0.4}$ & $78_{-18}^{+27}$ \\
J1042+1646 & SDSS J104244.23+164656.1 & 0.975 & 8.18 & BALQ & 47.12 & 9.04 & 0.93 & $46.9_{-0.1}^{+0.1}$ & $15_{-8}^{+8}, 800_{-200}^{+300}, 840_{-300}^{+500}$ \\
J1123+0137 & SDSS J112320.73+013747.4 & 1.475 & 8.68 & mini-BALQ & 47.73 & 9.82 & 0.63 & $44.9_{-0.2}^{+0.3}$ & $<22^{+37}, 340_{-190}^{+370}, 760_{-320}^{+440}, 1180_{-290}^{+430}$ \\
J1135+1615 & SDSS J113512.68+161550.6 & 2.006 & 8.58 & BALQ & 46.98 & 9.10 & 0.58 & $< 45.4$ & $<40^{+10}$ \\
J1209+1036 & SDSS J120924.07+103612.0 & 0.395 & 5.50 & mini-BALQ & 46.32 & 9.41 & 0.06 & $42.8_{-0.2}^{+0.3}$ & $500_{-110}^{+100}$ \\
J1329+5405 & SDSS J132957.14+540505.9 & 0.948 & 8.12 & BALQ & 46.96 & 9.13 & 0.52 & $> 43.4$ & $0.2_{-0.1}-0.4^{+17.4}$ \\
J1512+1119 & SDSS J151249.29+111929.3 & 2.113 & 8.52 & BALQ & 47.50 & 9.21 & 1.51 & $> 42.4$ & $10_{-1}-300^{+15}, >3000_{-150}$ \\
\hline
J0755+2306 & SDSS J075514.58+230607.1 & 0.853 & 7.88 & BALQ & 46.80 & 9.16 & 0.33 & $> 46.1$ & $270_{-90}^{+100}, 1600_{-1100}^{+2000}$ \\
J0936+2005 & SDSS J093602.10+200542.9 & 1.181 & 8.50 & mini-BALQ & 47.05 & 9.24 & 0.50 & $45.9_{-0.3}^{+0.2}$ & $14_{-4}^{+9}, 77_{-22}^{+40}, 150_{-50}^{+50}$ \\
J0941+1331 & SDSS J094111.11+133131.1 & 2.809 & 8.04 & BALQ & 47.26 & 9.54 & 0.40 & $44.6_{-0.1}^{+0.1}$ & $200_{-60}^{+40}$ \\
J1111+1437 & SDSS J111110.14+143757.0 & 2.162 & 8.50 & mini-BALQ & 47.06 & 9.14 & 0.65 & $43.8_{-0.1}^{+0.1}$ & $880_{-260}^{+210}$ \\
\hline
\end{tabular}
\begin{flushleft}
\vspace{-0.3cm}
\tablecomments{Columns:
(1) short name,
(2) full name,
(3) redshift,
(4) physical scale,
(5) BALQ or mini-BALQ,
(6) the logarithm scale of the bolometric luminosity,
(7) the logarithm scale of the BH mass based on the H$\beta$ or Mg\,II emission line,
(8) the Eddington ratio,
(9) the logarithm scale of the kinetic power,
(10) the outflow sizes of the multiple velocity components.
The first nine objects were detected in the observations, and the last four objects were not.
The bolometric luminosity and BH mass are obtained from \cite{Vestergaard2006,Vestergaard2009,Shen2011,Muzahid2012,Arav2013}, and the Eddington ratio is calculated based on them.
Both the $M_{\rm BH}$ and the $L/L_{\rm Edd}$ have a typical uncertainty of about $\pm$0.5\,dex.
The kinetic power and outflow sizes are taken from \cite{Miller2020,Byun2022} with 1$\sigma$ uncertainties.}
\end{flushleft}
\end{table}
\end{turnpage}

\begin{table}[ht!]
\caption{The radio coordinates and source sizes at 5.5 and 9.0\,GHz in the VLA A configuration observations.}
\label{size}
\centering
\footnotesize
\begin{tabular}{lcccccccc}
\hline\hline
Name & $\nu$ & R.A. & Dec. & $\theta_{\rm maj}$ & $\theta_{\rm min}$ & $d_{\rm maj}$ & $d_{\rm min}$ & PA \\
& (GHz) & (hh:mm:ss) & (dd:mm:ss) & (arcsec) & (arcsec) & (kpc) & (kpc) & (degree) \\
(1) & (2) & (3) & (4) & (5) & (6) & (7) & (8) & (9) \\
\hline
J0240$-$1851 & 5.5 & 02:40:32.584 & $-$18:51:51.360 & $<$0.52 & $<$0.20 & $<$3.66 & $<$1.41 & \\
& 9.0 & 02:40:32.583 & $-$18:51:51.369 & 0.44 & 0.35 & 3.09 & 2.46 & 26 \\
J0242+0049 & 5.5 & 02:42:21.877 & +00:49:12.624 & $<$0.28 & $<$0.20 & $<$2.40 & $<$1.71 & \\
& 9.0 & 02:42:21.872 & +00:49:12.694 & $<$0.17 & $<$0.12 & $<$1.46 & $<$1.03 & \\
J0831+0354 & 5.5 & 08:31:26.155 & +03:54:08.096 & 0.50 & $<$0.20 & 4.28 & $<$1.71 & 128 \\
& 9.0 & 08:31:26.155 & +03:54:08.112 & $<$0.21 & $<$0.12 & $<$1.80 & $<$1.03 & \\
J1042+1646 & 5.5 & 10:42:44.240 & +16:46:56.120 & $<$0.30 & $<$0.21 & $<$2.45 & $<$1.72 & \\
& 9.0 & 10:42:44.236 & +16:46:56.104 & $<$0.21 & $<$0.13 & $<$1.72 & $<$1.06 & \\
J1123+0137 & 5.5 & 11:23:20.721 & +01:37:47.521 & $<$0.24 & $<$0.18 & $<$2.08 & $<$1.56 & \\
& 9.0 & 11:23:20.725 & +01:37:47.494 & $<$0.14 & $<$0.11 & $<$1.22 & $<$0.95 & \\
J1135+1615 & 5.5 & 11:35:12.691 & +16:15:50.548 & 0.34 & $<$0.19 & 2.92 & $<$1.63 & 166 \\
& 9.0 & 11:35:12.693 & +16:15:50.576 & 0.30 & $<$0.12 & 2.57 & $<$1.03 & 127 \\
J1209+1036 & 5.5 & 12:09:24.084 & +10:36:12.044 & 0.25 & $<$0.18 & 1.38 & $<$0.99 & 32 \\
& 9.0 & 12:09:24.084 & +10:36:12.018 & 0.19 & 0.17 & 1.05 & 0.94 & 153 \\
J1329+5405 & 5.5 & 13:29:57.150 & +54:05:06.012 & $<$0.22 & $<$0.18 & $<$1.79 & $<$1.46 & \\
& 9.0 & 13:29:57.151 & +54:05:05.972 & $<$0.13 & $<$0.12 & $<$1.06 & $<$0.97 & \\
J1512+1119 & 5.5 & 15:12:49.295 & +11:19:29.342 & $<$0.26 & $<$0.20 & $<$2.22 & $<$1.70 & 134 \\
& 9.0 & 15:12:49.297 & +11:19:29.406 & $<$0.18 & $<$0.12 & $<$1.53 & $<$1.02 & \\
\hline
\end{tabular}
\begin{flushleft}
\vspace{-0.3cm}
\tablecomments{Columns:
(1) short name,
(2) frequency,
(3) right ascension,
(4) declination,
(5) the de-convolved major FWHM of the source,
(6) the de-convolved minor FWHM of the source,
(7) the physical size of the major axis of the source,
(8) the physical size of the minor axis of the source,
(9) the de-convolved position angle of the source.}
\end{flushleft}
\end{table}

\begin{table}[ht!]
\caption{The total and peak flux densities, and the background noise in the full-array maps and the tapered maps at 5.5 and 9.0\,GHz in the VLA A configuration observations. The tapered maps have a $uv$ range of $\sim$ 20--660\,k$\lambda$ in both bands, corresponding to an angular coverage of $\sim$ 0.4\arcsec--13\arcsec.}
\label{flux}
\centering
\footnotesize
\begin{tabular}{lccccccc}
\hline\hline
Name & Frequency & \multicolumn{3}{c}{Full-array maps} & \multicolumn{3}{c}{Tapered maps} \\
& $\nu$ & $F_{\rm total}$ & $F_{\rm peak}$ & RMS & $F_{\rm total}$ & $F_{\rm peak}$ & RMS \\
& (GHz) & (mJy) & (mJy\,beam$^{-1}$) & (mJy\,beam$^{-1}$) & (mJy) & (mJy\,beam$^{-1}$) & (mJy\,beam$^{-1}$) \\
(1) & (2) & (3) & (4) & (5) & (6) & (7) & (8) \\
\hline
\multirow{2}{*}{J0240$-$1851} & 5.5 & 0.353 $\pm$ 0.012 & 0.351 $\pm$ 0.006 & 0.058 & 0.964 $\pm$ 0.118 $^\dagger$ & 0.421 $\pm$ 0.037 & 0.075 \\
& 9.0 & 0.791 $\pm$ 0.082 & 0.360 $\pm$ 0.027 & 0.066 & 0.802 $\pm$ 0.094 & 0.354 $\pm$ 0.030 & 0.049 \\
\hline
\multirow{2}{*}{J0242+0049} & 5.5 & 0.045 $\pm$ 0.004 & 0.047 $\pm$ 0.002 & 0.008 & 0.043 $\pm$ 0.004 & 0.046 $\pm$ 0.002 & 0.008 \\
& 9.0 & 0.033 $\pm$ 0.002 & 0.024 $\pm$ 0.001 & 0.007 & 0.047 $\pm$ 0.003 & 0.024 $\pm$ 0.001 & 0.007 \\
\hline
\multirow{2}{*}{J0831+0354} & 5.5 & 0.111 $\pm$ 0.014 & 0.076 $\pm$ 0.006 & 0.009 & 0.109 $\pm$ 0.014 & 0.076 $\pm$ 0.006 & 0.009 \\
& 9.0 & 0.032 $\pm$ 0.006 & 0.039 $\pm$ 0.004 & 0.009 & 0.036 $\pm$ 0.004 & 0.043 $\pm$ 0.002 & 0.010 \\
\hline
\multirow{2}{*}{J1042+1646} & 5.5 & 0.069 $\pm$ 0.005 & 0.063 $\pm$ 0.003 & 0.010 & 0.067 $\pm$ 0.005 & 0.063 $\pm$ 0.003 & 0.010 \\
& 9.0 & 0.045 $\pm$ 0.007 & 0.058 $\pm$ 0.005 & 0.010 & 0.046 $\pm$ 0.010 & 0.059 $\pm$ 0.006 & 0.010 \\
\hline
\multirow{2}{*}{J1123+0137} & 5.5 & 0.043 $\pm$ 0.005 & 0.042 $\pm$ 0.003 & 0.012 & 0.047 $\pm$ 0.004 & 0.043 $\pm$ 0.002 & 0.011 \\
& 9.0 & 0.035 $\pm$ 0.004 & 0.041 $\pm$ 0.003 & 0.011 & 0.030 $\pm$ 0.003 & 0.045 $\pm$ 0.002 & 0.011 \\
\hline
\multirow{2}{*}{J1135+1615} & 5.5 & 0.140 $\pm$ 0.013 & 0.102 $\pm$ 0.006 & 0.010 & 0.142 $\pm$ 0.014 & 0.102 $\pm$ 0.006 & 0.010 \\
& 9.0 & 0.067 $\pm$ 0.009 & 0.042 $\pm$ 0.004 & 0.012 & 0.062 $\pm$ 0.006 & 0.045 $\pm$ 0.003 & 0.013 \\
\hline
\multirow{3}{*}{J1209+1036} & 5.5 & 0.475 $\pm$ 0.024 & 0.394 $\pm$ 0.012 & 0.011 & 0.464 $\pm$ 0.024 & 0.390 $\pm$ 0.012 & 0.011 \\
& 5.5 & 0.163 $\pm$ 0.025 & 0.062 $\pm$ 0.007 & 0.011 & 0.158 $\pm$ 0.027 & 0.055 $\pm$ 0.007 & 0.011 \\
& 9.0 & 0.276 $\pm$ 0.026 & 0.182 $\pm$ 0.011 & 0.012 & 0.285 $\pm$ 0.025 & 0.199 $\pm$ 0.011 & 0.012 \\
\hline
\multirow{2}{*}{J1329+5405} & 5.5 & 0.115 $\pm$ 0.015 & 0.098 $\pm$ 0.008 & 0.011 & 0.118 $\pm$ 0.015 & 0.097 $\pm$ 0.008 & 0.011 \\
& 9.0 & 0.067 $\pm$ 0.011 & 0.049 $\pm$ 0.005 & 0.012 & 0.108 $\pm$ 0.010 & 0.046 $\pm$ 0.003 & 0.012 \\
\hline
\multirow{2}{*}{J1512+1119} & 5.5 & 0.109 $\pm$ 0.010 & 0.090 $\pm$ 0.005 & 0.014 & 0.102 $\pm$ 0.008 & 0.089 $\pm$ 0.004 & 0.014 \\
& 9.0 & 0.135 $\pm$ 0.013 & 0.129 $\pm$ 0.007 & 0.010 & 0.131 $\pm$ 0.014 & 0.133 $\pm$ 0.008 & 0.011 \\
\hline
\multirow{2}{*}{J0755+2306} & 5.5 & & $<$0.036 & 0.012 & & & \\
& 9.0 & & $<$0.030 & 0.010 & & & \\
\hline
\multirow{2}{*}{J0936+2005} & 5.5 & & $<$0.036 & 0.012 & & & \\
& 9.0 & & $<$0.030 & 0.010 & & & \\
\hline
\multirow{2}{*}{J0941+1331} & 5.5 & & $<$0.036 & 0.012 & & & \\
& 9.0 & & $<$0.030 & 0.010 & & & \\
\hline
\multirow{2}{*}{J1111+1437} & 5.5 & & $<$0.036 & 0.012 & & & \\
& 9.0 & & $<$0.033 & 0.011 & & & \\
\hline
\end{tabular}
\begin{flushleft}
\vspace{-0.3cm}
\tablecomments{Columns:
(1) short name,
(2) frequency,
(3) the total flux density of the full-array map,
(4) the peak flux density of the full-array map,
(5) the background noise of the full-array map,
(6) the total flux density of the tapered map,
(7) the peak flux density of the tapered map,
(8) the background noise of the tapered map.
The symbol $\dagger$ marks an inconsistent flux due to systematic uncertainties, e.g., a higher background noise level.}
\end{flushleft}
\end{table}

\begin{table}[ht!]
\caption{The observed 5.5--9.0\,GHz spectral slope, the radio loudness, and the rest-frame 5.5 and 9.0\,GHz radio luminosities of the sample.}
\label{result}
\centering
\footnotesize
\begin{tabular}{lrrrr}
\hline\hline
Name & $\alpha_{5.5-9.0}$ & $R$ & $\log L_{\rm 5.5GHz}$ & $\log L_{\rm 9.0GHz}$ \\
& & & (erg\,s$^{-1}$) & (erg\,s$^{-1}$) \\
(1) & (2) & (3) & (4) & (5) \\
\hline
J0240$-$1851 & $-$0.35 $\pm$ 0.25 & 0.18 & 40.59 & 41.15 \\
J0242+0049 & $-$1.32 $\pm$ 0.12 & 0.61 & 41.54 & 41.60 \\
J0831+0354 & $-$1.16 $\pm$ 0.19 & 2.16 & 41.85 & 41.61 \\
J1042+1646 & $-$0.13 $\pm$ 0.23 & 0.06 & 40.31 & 40.45 \\
J1123+0137 & 0.09 $\pm$ 0.13 & 0.02 & 40.47 & 40.67 \\
J1135+1615 & $-$1.66 $\pm$ 0.18 & 4.55 & 42.14 & 42.04 \\
J1209+1036 C1 & $-$1.37 $\pm$ 0.13 & 0.97 & 40.35 & 40.33 \\
J1209+1036 C2 & $< -0.86 \pm 0.26$ & & & \\
J1329+5405 & $-$1.51 $\pm$ 0.21 & 0.45 & 40.90 & 40.88 \\
J1512+1119 & 0.82 $\pm$ 0.15 & 0.07 & 40.89 & 41.20 \\
\hline
J0755+2306 & & $<$0.09 & $<$39.98 & $<$40.11 \\
J0936+2005 & & $<$0.08 & $<$40.36 & $<$40.50 \\
J0941+1331 & & $<$0.59 & $<$41.41 & $<$41.54 \\
J1111+1437 & & $<$0.17 & $<$41.09 & $<$41.27 \\
\hline
\end{tabular}
\begin{flushleft}
\vspace{-0.3cm}
\tablecomments{Columns:
(1) short name,
(2) the observed 5.5--9.0\,GHz spectral slope,
(3) the radio loudness,
(4) the logarithm scale of the rest-frame 5.5\,GHz radio luminosity,
(5) the logarithm scale of the rest-frame 9.0\,GHz radio luminosity.}
\end{flushleft}
\end{table}

\end{document}